# Viscoelastic properties of green wood across the grain measured by harmonic tests in the range of 0°C to 95°C. Hardwood vs. softwood and normal wood vs. reaction wood.

Vincent Placet, Joëlle Passard, Patrick Perré

Laboratoire d'Etude et de Recherche sur le MAtériau Bois

LERMAB-UMR INRA/ENGREF/UHP 1093

ENGREF 14, rue Girardet 54042 Nancy cedex, France

Tel: 33 (0)3 83 39 68 92

Fax: 33 (0)3 83 39 68 47

E-mail: placet@nancy-engref.inra.fr

## Abstract

The viscoelastic properties of wood have been investigated with a dynamic mechanical analyser (DMA) specifically conceived for wooden materials, the WAVE$^T$ device (environmental vibration analyser for wood). Measurements were carried out on four wood species in the temperature range of 0°C to 100°C at frequencies varying between 5 mHz and 10 Hz. Wood samples were tested in water-saturated conditions, in radial and tangential directions. As expected, the radial direction always revealed a higher storage modulus than the tangential direction. Great differences were also observed in the loss factor. The tanδ peak and the internal friction are higher in tangential direction than in radial direction. This behaviour is attributed to the fact that anatomical elements act depending on the direction. Viscoelastic behaviour of reaction wood differs from that of normal or opposite wood. Compression wood of spruce, which has higher lignin content, is denser and stiffer in transverse directions than normal wood, and has lower softening temperature (Tg). In tension wood, the G-layer is weakly attached to the rest of the wall layers. This may explain why the storage modulus and the softening temperature of tension wood are lower than those for the opposite wood.

In this work, we also point out that the time-temperature equivalence fits only around the transition region, i.e. between Tg and Tg + 30°C. Apart from these regions, the wood response combines the effect of all constitutive polymers, so that the equivalence is not valid anymore.

## 1. Introduction

The softening transition of polymers is an important parameter which is considered since a long time in numerous processes of manufacture and polymer transformation. This concept has also been applied for food processing (Genin and René 1995) and studying wooden materials. It is well known that the softening of wet wood is in the range of 50°C to 100°C which is closely related to the lignin transition. Hemicelluloses and amorphous cellulose in wet wood have lower softening temperatures than those of lignins (Olsson and Salmén 1992; Lenth and Kamke 2001). Moreover, this property depends on the wood species (Olsson and Salmen 1992; Dwianto *et al*. 1998; Takahashi *et al*. 1998; Hamdan *et al*. 2000).

Juvenile wood (Lenth 1999) and reaction wood (Olsson and Salmén 1992) differ in their softening behaviour. The lignin content and lignin structure exert a great influence on softening. Lignin is a branched polymer made up of phenylpropane units. The softening temperature of wood, noted Tg, increases with the degree of cross-linking of the lignin (Olsson and Salmén 1992). Methoxyl groups hamper crosslinking of the aromatic units. Their presence in abundance in typical hardwoods, and this fact results in a more flexible network with lower Tg. As water acts as a plasticizing agent for wood, the moisture content also affects the Tg (Goring 1963; Ranta-Maunus 1973; Irvine 1984; Ebrahimzadeh and Kubat 1993; Navi *et al.* 2000; Obataya *et al.* 1998, 2001). Water also give rise to hydrogen bonds with methoxyl groups (Birkinshaw 1993; Obataya *et al.* 2003). The replacement of hydrogen bonding within the polymer network of the cell wall by bonds between water-lignin, water-hemicelluloses, and water-paracristalline cellulose enhances the flexibility of the polymer network. Numerous methods are available for characterisation of the effect of temperature or moisture on the Tg of the cell wall. There are publications on the softening behaviour of isolated polymers of wood (Goring 1963; Irvine 1984; Olsson and Salmén 1997) and of solid wood (Olsson and Salmén 1992; Lenth and Kamke 2001; Obataya *et al.* 2003; Passard and Perré 2005).

A method based on harmonic tests, the dynamic mechanical analysis (DMA), is a quite common method to decouple thermal activation from the time effect. In DMA, a harmonic force (or a displacement) is applied to the sample and its displacement (or the force) is measured. A phase difference can be observed between the resultant response and the applied stimulus. This is dependent of the properties of the polymer investigated. Thus, the storage modulus E' is defined as the in-phase or elastic response, proportional to the recoverable or stored energy. The loss modulus E'' is the imaginary or viscous response, proportional to the irrecoverable or dissipated energy. The loss factor, tan$\delta$ is the ratio E''/E'; in words: it represents the ratio of the dissipated energy to the stored energy. This is one of the key parameters in DMA because its value varies dramatically in the course of the polymer transition.

Many researchers have measured dynamic mechanical properties of wood in the longitudinal direction or across the grain (Kelley *et al.* 1987; Olsson and Salmén 1992; Obataya *et al.* 2003; Backman and Lindberg 2001). Backman and Lindberg (2001) remark that few works are devoted to the DMA of the radial and tangential directions. Accordingly, our knowledge about the softening of these anatomical directions is poor.

To close this gap, in this work DMA results are presented which also includes the radial and tangential direction over wide ranges of temperature and frequency. The experiments were performed with a special apparatus specifically conceived for soaked samples. This experimental device is described in detail by Placet (2006) and Placet *et al.* (2007). The DMA behaviour of various species and reaction woods will be described. The time-temperature equivalence should also be illuminated.

## 2. Materials and methods

### 2.1 Experimental apparatus

Dynamic mechanical properties of green wood were measured between 0°C to 95°C and at frequencies from 5 mHz to 10 Hz. It is important to note that large differences may be observed between data obtained by different commercial DMA equipments. The interlaboratory reproducibility of results obtained by the same or similar instruments is also poor (Hagen *et al.* 1994). These differences may often be explained by problems related to mechanical inertia, specimen geometry and size, clamping effects, thermal lag, inertia corrections, just to mention a few. To avoid these inadequacies and to address several wood specificities, such as anisotropy and hygroscopicity, a new experimental device, the WAVE$^T$ (environmental vibration analyser for wood), was used here (Figure 1). The WAVE$^T$ is a custom apparatus, developed at the Wood Research Laboratory (LERMAB in France). A detailed description of this device and the modelling approach developed to analyse the raw data can be found in Placet (2006) and Placet *et al.* (2007). However, its most important specifications are summarised here.

The WAVE$^T$ is able to determine the viscoelastic properties (stiffness and damping properties) of water-saturated samples between 5°C to 95°C, at frequencies $5.10^{-3}$ to 10 Hz and at stress levels 0.01 to 4 MPa. In this forced non-resonance technique, specimens are tested in a bending configuration based on a rigorous single cantilever mode. To keep samples water-saturated during the experimentation and avoid any mecanosorptive effects due to temperature gradient, the samples are placed in a temperature-controlled water bath throughout the test. Moreover, contrary to most commercial instruments, the force applied to the sample is not calculated from knowledge of the input signal to the driver but actually measured by a miniature load cell placed between the sample and the driver. As additional advantage, this load cell allows the applied force to be controlled by a closed-loop regulation. To ensure rigorous mechanical measurements, the sample holder has been conceived with great care. A specific articulated clamp ensures that a pure vertical force is applied without any momentum to the sample. By this way, a pure single cantilever configuration is obtained. Finally, in order to avoid any disturbance at the clamp area, the deflection measurement is dissociated from the other part of the system.

Besides, great care was taken to collect, treat, and analyse raw data. For example, a comprehensive model of the whole device is implemented in the control software to account for the inertia of the moving part and correct its effects in the measured parameters.

At each frequency and each temperature level, the viscoelastic parameters (storage modulus E', loss modulus E'' and loss factor tan$\delta$) are determined taking into consideration of the macroscopic dimensions of the wood specimens at 20°C, the measured force and deflection, and the phase difference. The latter is a key parameter and has to be determined with great accuracy. This is why it is identified by an inverse method by means of an objective function which defines the difference between the raw data and perfect harmonic functions over several periods.

There are many ways to identify the value of Tg based on the dynamic mechanical measurements, such as the peak value of the loss tangent, the peak value of the loss modulus, and the onset of the drop of the storage modulus. However, one has to be aware of the fact that the obtained Tg value depends on the method (Hagen *et al.*1994). Even though we do know that the tan$\delta$ peak is not directly tied to the chain mobility, we chose this method here to define the softening temperature. The peak value of the loss factor is the only parameter always clearly observable in our harmonic tests. This finding probably explains why many authors applied the same method (Salmén 1984; Olsson & Salmén 1992; Kelley *et al*. 1987).

## 2.2   Wood material

The hardwoods oak (*Quercus sessiliflora*), beech (*Fagus sylvatica*), and poplar (*Populus sp.*), and as a softwood spruce (*Picea abies*) were tested. For each wood species, two series of samples were cut from the heartwood part of a green log, one for radial tests and the other for tangential tests. For each set, samples are successive sections of a board perfectly parallel to the longitudinal direction of the trunk. This precaution allows the same annual growth rings to be tested, and to obtain a set of matched samples with very similar properties (Placet *et al.* 2007). The sample section is 5 x 10 mm², its length 100 mm. Figure 2 depicts a schematic example of wood sampling. Note that in tangential direction, the samples have an annual ring structure which is not symmetrical.

Once cut, the samples are immediately soaked in water in an airtight box and put in a refrigerator to limit fungal attacks. Then, samples are tested within a few days. Oak and beech trees were harvested nearby the Forest School of Nancy (ENGREF). The poplar tree was a young tree coming from an experimental stand (Unit of Forest Improvement, Genetics and Physiology, INRA Orléans). Planted in 1996, it was artificially bent from 1998 to induce tension wood formation. Its tension wood was easily visible by the fuzzy grain due to gelatinous fibers. Microscopic views prepared with an ESEM electron microscope confirmed the presence of G layers, easily separated from the rest of the secondary cell wall. The "opposite wood" is defined as the wood produced on the other side of the tension wood.

Spruce was taken from the region of Nancy, France. This tree developed compression wood after the historical storms, named "Lothar" and "Martin", which damaged large forest areas in France in 1999. Microscopic images confirmed the absence of compression wood before this date. Therefore, we cut samples in compression wood and "normal" wood. The limited width of the compression wood region did not permit us the preparation of radial samples.

For each wood species, anatomical direction and wood type, several matched samples were tested. The following results are presented for one of these matched samples. The repeatability of the measurements was tested on at least two samples.

## 3. Results and discussion

### 3.1 Viscoelastic behaviour of green wood across the grain

*Normal wood*

Figure 3 exhibits the storage modulus and the loss factor as function of temperature at different frequencies, for oak, beech, and spruce in radial direction. These curves reveal the typical characteristics of the viscoelastic behaviour of wood. The storage modulus decreases with increasing temperature and a softening transition appears in the temperature range 70 to 100°C, corresponding to lignin relaxation. The storage modulus curves are perfectly staggered. Actually, as for all viscoelastic solid materials, the storage modulus of wood increases with frequency. Moreover, the maximum value of the loss factor increases with frequency, which attests that the transition temperature shifts to higher values as the frequency increases.

Beyond theses common trends, some differences appear between species. For a frequency of 1 Hz and a temperature of 20°C, the radial storage modulus is 740 MPa for beech, 580 MPa for oak, and 360 MPa for spruce. Of course, this inter-species variability has to be assessed in comparison to intra-species variability. In addition, the softening temperature is slightly higher and the internal friction is lower for beech than for oak (at 1 Hz for oak $T_g$ = 78°C with a $\tan\delta$ of 0,135, for beech $T_g$ = 83°C with a $\tan\delta$ of 0,108). Over this temperature range (5°C – 95°C), the curve deflection due to the transition zone just emerges for spruce at the highest temperature values at 0.1 Hz. Unfortunately, due to experimental problems, the curve at 0.1 Hz, which may have confirmed this trend is neither available on this spruce sample nor on the matched sample. Anyway, these results show that the softening of softwood actually occurs at higher temperatures than that for hardwood (Olsson and Salmén 1992).

For a precisely capture the softening temperature of softwood, two solutions are conceivable: to perform tests above 100°C or the application of lower frequencies. Tests with soaked samples over 100°C cannot be performed with the present WAVE$^T$ version which works at atmospheric pressure. (A second WAVE$^T$ device designed for 5 bar pressure is suited to temperatures above 100°C and the results obtained will be published later.) In the present paper we applied the lower frequency values between 0.1 Hz and 0.005 Hz for characterization of spruce. This technique made the loss factor peak visible. Accordingly, the softening temperature of spruce in radial direction is equal to 70°C at 5 mHz. The drawback of this type of experimentation is its long duration, which may induce thermal degradation. Tests performed on beech samples showed that the viscoelastic properties of soaked sample are significantly affected even at moderate temperature levels: (a) a decrease of more than 3% was measured after 3 h above 80°C for the storage modulus and (b) a decrease of about 15% was measured after 8 h at 93°C. More information concerning the effect of the hygrothermal treatment on the viscoelastic properties of wood can be found in Placet (2006).

At the left hand part of the curves, based on the trend observed near 0°C a second peak seems to be possible, which may be attributed to softening of hemicelluloses. Consistently with results of Backman and Lindberg (2001), the storage modulus depicts also a clear difference between radial and tangential directions, whatever frequencies, temperatures, and species are chosen (Figure 4). For the trees investigated in our study, beech is stiffer than oak, but the ratio $E'_R/E'_T$ is of the same magnitude for the two species: 1.37 for beech and 1.41 for oak. New finding is that the internal friction and the tanδ peak are higher in tangential than in radial direction. For both species, the peak value of the loss factor occurs at 5 to 10°C higher temperature in tangential than in radial direction.

*Reaction wood*

Reaction wood and normal wood show different behaviour (Table 1). Whatever the material direction, tension wood of poplar is less stiff than its opposite wood, and the softening temperature of tension wood is lower than that of opposite wood. For spruce, the basic density almost doubles for compression wood, and the storage modulus is almost five times higher than that of normal wood in the transverse plane. However, the softening temperature is lower for compression wood (Figure 5).

The viscoelastic properties obtained for these four species are summarized in Tables 1 and 2. Whatever the wood species, the storage modulus determined at 1 Hz decreases of about 70% in radial direction and of about 80% in tangential direction when the temperature increases from 5°C to 90°C (Table 2).

### 3.2 Discussion about the differences found in wood behaviour

Our measurements pointed out many differences with regard to the rheological behaviour of woods depending on the species, the type of wood, and the anatomical directions. The anatomical and molecular features have the highest influence. To express it more precisely: the molecular structure of the essential wood components and their supramolecular architecture within the cell wall influence the softening properties essentially.

Water also acts as a plasticizer. In water-saturated wood, the interactions between macromolecules in the cell wall are weaker. Furthermore, the relaxation of the hemicelluloses in wet wood is shifted just below 0°C (Kelley *et al.* 1987); only the softening of lignin is visible between 0℃ to 100℃. In this temperature range, the wood behaviour reflects to a large extent the properties of lignin.

In lignin, the phenylpropane units are linked together to a branched and partly crosslinked polymer network. The chains mobility is restricted in the interpenetrating network of the cell wall. The increased amount of hydrogen bonds under wet conditions facilitates the flexibility within the micrcofibrils and the cell wall layers. The number of methoxyl groups (OMe groups) plays an important role with this regards. A syringil unit (S) has two and a guaiacyl unit (G) has one of this functional group. The ratio S/G units is significantly higher in hardwoods than in softwoods (Olsson & Salmén 1992). This is the reason why the former is less cross-linked and has a lower softening point than the latter.

Pilate *et al.* (2004) stated that the lignin in cell wall of tension wood differs from that of normal wood, particularly due to the elevated ratio of S/G units. Consequently, the softening temperature of tension wood is lower. It is also well known that the lignin of the compression wood of spruce is enriched in *para*-hydroxyphenylpropane units (without OMe groups); a fact which opens the possibility for cross-linking reactions and which could lead to higher softening temperatures. However, in our experiments the softening temperature of compression wood is decreased. These results demonstrate that the interpretation of the DMA results cannot be done on the molecular level alone and based on simplified models. The cell wall is more than a mixture of polymers. The physical behaviour of wood is influenced of the supramolecular architecture of the cell wall. The internal friction level within the cell wall is decisive in this context. The more material is involved in the softening mechanism, the higher is the tanδ. The microfibril angle should also be taken into consideration to explain the level of the internal friction.

Biochemical tests were carried out on our samples at INRA Reims following the protocol proposed in the papers of Liyama and Wallis (1990) and Beaugrand *et al.* (2004). As expected, tension wood of poplar has lower lignin content than its opposite wood (13.4% vs. 19.3%), and compression wood of spruce is richer in lignin than normal wood (40.8% vs. 33.7%).

Measurement with WAVE$^T$ show that the enrichment in lignin for compression wood of spruce results in an increased loss factor ($\tan\delta_{max}$ of compression wood is near 1.4 times higher than $\tan\delta_{max}$ of normal wood). But, in the tension wood of poplar, despite its lower lignin amount, we measured an important increasing of the internal friction. One of the possible interpretations: Though the average lignin content of the tension wood is low due to the broad G-layer, which consists of cellulose, rest of the wall may be very rich in lignin. The latter is probably decisive for the softening.

The average composition of wood does not explain the differences in the supramolecular structures in radial and tangential directions. As the tanδ value is systematically higher in tangential direction, it can be concluded that the internal friction and the proportion of material being involved in this process are larger in this direction.

Another aspect is also important for interpretation: The wood tissue is more regular in radial direction compared to tangential direction. Cells are aligned in radial direction and organised in staggered rows in tangential direction. In bending tests, the staggered random cells arrangement behaves differently than a well organised row. In radial direction, the cell walls are loaded mainly in pure traction and compression whereas in tangential direction, cell walls are rather loaded in flexion (Figure 6). The shear stress due to flexion induces a cellular sliding which most probably occurs in the middle lamella. Now, it is well known that the lignin distribution along the double cell wall is heterogeneous (Yoshizawa *et al.* 1999; Salmén 2004). According to Mark (1967), the lignin content is nearly two times higher in the middle lamella than in other wall layers. If we consider that the lignin rich middle lamella is the main element submitted to shear strain in tangential direction, the finding of the high tan$\delta$ value is in tangential direction is more understandable.

In radial direction, the rigidity is enhanced because of the presence of ray cells. They are stiffening elements of the composite matrix of wood. But for certain species the ray cells walls are thin and the average micro fibril angle (MFA) is large, and so their stiffening effect is not sufficient to explain the large differences observed. Another explanation of the enhanced rigidity in radial direction lies in the preferential organisation of cells along this direction, due to the cell lines produced by the same mother cell in the cambial zone.

The rigidity of reaction wood differs from normal wood. For compression wood, the increased MFA adds to the effect of basic density in the transverse directions. For tension wood, the G-layer is detached from the other wall layers and does not take part in the stiffness increasing of wood in the transverse plane. Consequently, the part of wall cell responsible for the wood rigidity is thinner than in opposite wood. The low MFA in tension wood also participates to a low rigidity value across the grain.

## 3.3 Time-temperature equivalence

In wood rheology, the time-temperature equivalence is an important concept discussed for a long time. This term was widely used by numerous authors (Salmén 1984; Geneveaux 1989; Bardet 2001; Lenth and Kamke 2001; Le Govic 1992; Perré and Aguiar 1999).

The thermal activation can be expressed by the WLF law (Williams, Landel, and Ferry) and the Arrhenius law. Both laws are valid over a certain range of temperature. Because the WLF law originates from the free volume theory, it is valid only for temperature values above the softening temperature (Williams *et al.* 1955). Actually, molecular motion at the relaxation level needs volumetric expansion. The first step of the WLF analysis consists in building the master curve based on raw data. To do this, the curves storage modulus vs. frequency at various temperatures are shifted along the x-axis to build up a single curve. The shift value of each curve obtained at temperature T from the reference curve obtained at temperature $T_0$ defines the shift factor $a_{T,T_0}$. In Figure 7-a master curves are depicted which were obtained for beech and oak in radial and tangential directions. These curves have been constructed from the measurements performed by the WAVE$^T$ apparatus at various frequencies and temperatures, with a temperature reference of 75°C. The WLF law says that the shift factor depends on temperature as follows:

$$\log a_{T,T_0} = -\frac{c_1^0(T-T_0)}{c_2^0+T-T_0} \qquad (1)$$

where $a_{T,T_0}$ is the shift factor, $T_0$ the reference temperature (K), $c_1^0$ and $c_1^0$ two WLF constants. If the reference temperature is $T_g$, the two WLF constants are noted $C_1$ and $C_2$, and are supposed to be universal.

Figure 7-b shows that there is a good agreement between the experimental data and the prediction of the WLF equation from about 55°C (in the case of beech in radial direction). This temperature corresponds to the softening temperature of wood at very low frequencies (Salmén 1984). Table 3 synthesises the softening temperature determined with this method for beech and oak samples in radial and tangential directions. It appears that the softening temperature is lower in radial

direction than in tangential direction; the difference being of about 5°C for both species. This result, determined by the WLF law, confirmed the tendency emphasised in paragraph 3.1 based on the peak of tanδ.

The WLF constants ($C_1$ and $C_2$) were determined from the linear part of the curve $(T-T_0)/\log a_T$ versus $(T-T_0)$. Table 3 summarises the WLF parameters (constants $C_1$, $C_2$ and the activation energy), determined for a temperature reference corresponding to the softening temperature of these wood species at very low frequencies. The values of these constants are relatively similar. Figure 7-b depicts the evolution of $\log a_T$ as a function of $(T-T_0)$ for a beech sample in radial direction. For classical polymers, the WLF law is considered to be valid between $T_g$ and $T_g$+100K, but Salmén (1984) found a reduced range for wood, from $T_g$ to $T_g$+55K. In this work, we obtained an even more reduced range from $T_g$ to $T_g$+30K.

Below the softening temperature, the free volume is independent of temperature; the mobility variation with temperature can be determined following an Arrhenius law, expressed as follows:

$$\frac{d\ln(f)}{d\left(\frac{1}{T}\right)} = \frac{d\ln(a_{T,T_0})}{d\left(\frac{1}{T}\right)} = \frac{-\Delta H_a}{R} \tag{2}$$

where f is the frequency (Hz), T the temperature (K), $\Delta H_a$ the apparent activation energy (J.mol$^{-1}$) and R the gas constant (8.314 J.mol$^{-1}$K$^{-1}$).

Figure 8 depicts the Arrhenius plot. The $T_g$ value obtained from WLF data is also shown for comparison, with the low frequency set to near 10$^{-4}$ Hz (dotted lines in Figure 8). This frequency is just an approximation, previously suggested by Hagen *et al.* (1994). Actually, it is generally accepted that the softening temperature determined by DSC, dilatometric methods or WLF data is corresponding to very low frequencies, i.e. at about 10$^{-4}$ Hz. Therefore, the extrapolation of the Tg value determined from the maximum value of the loss factor is in good agreement with the WLF value, for the corresponding anatomical direction. These curves allow the apparent activation energy to be determined. The corresponding values are lower than those found by the WLF law (Table 3).

Cole-Cole diagrams represent a convenient way to synthesised viscoelastic properties of materials (Cole and Cole 1941, 1942). It permits the visualisation of the relaxation mechanisms. The viscoelastic data gathered with harmonic tests by WAVE$^T$ easily exhibit the lignin transition when plotted in the Cole-Cole plot (Figure 9-a). In this diagram, transitions are traduced by circle arcs. This main circle arc is broken by a second one, certainly due to the relaxation of hemicelluloses. This confirms the hypothesis expressed earlier concerning the possible existence of a second transition near 0°C corresponding to the softening of hemicelluloses. For this oak sample in radial direction, the time-temperature equivalence seems to be effective.

However, when data are plotted vs. frequency for different temperatures (Figure 9-b), the curves are no longer superimposed for high frequencies and low temperatures. Yet, in this thermo-temporal range, just a wide part of the lignin relaxation mechanism and the beginning of the relaxation of hemicelluloses is visible. The time-temperature equivalence can not be applied on the whole viscoelastic range, but seems to be valid within each transition state. Within this zone, the particular behaviour of one softening polymer prevails over the other constitutive polymers of wood. The finding confirms what Le Govic (1992) claimed more than one decade ago.

Rheological models may be easily represented in Cole-Cole diagrams, and particularly with thermoactivated Kelvin's elements (Passard and Perré 2005). In this representation, each Kelvin's element produces a circle arc whose centre is on the x-axis. However, experimental data of Figure 8 show that the circle arcs are flattened for wood. The loss modulus is about ten times lower than the storage modulus. There are two analog models for simulation of this complex behaviour: (1) The input of certain numbers of Kelvin's (or Maxwell) interacting elements (Bardet 2001) and (2) the input of parabolic elements (Huet, 1988).

Unfortunately, such models loose their physical meaning and are difficult to implement in a computational model. Constitutive models based on structural and molecular behaviour are much more promising with this regard.

## 4. Conclusion

We presented a set of wood viscoelastic properties collected by means of an improved dynamic mechanical analyser, conceived specially for wood at LERMAB. Among the numerous results presented in this paper for wet wood, three major points have to be emphasised:

1. Viscoelastic properties of wood are noticeably different in radial and tangential directions. Tangential direction has a lower rigidity, a higher softening temperature and higher internal friction than radial direction.
2. Reaction wood is also distinguished from normal and opposite wood. These differences can be explained by the anatomical elements activated during transitions, and by the particular bio-chemical composition of reaction wood.
3. Time-temperature equivalence can not be applied on the whole viscoelastic range, but is valid within each transition zone.

Our next project will be to investigate the behaviour of wet wood at temperature levels above 100°C, which requires mechanical tests to be carried out under pressure. We hope from these measurements to clarify the open questions of the time-temperature equivalence.

**Acknowledgements:**

We would like to gratefully acknowledge G.Pilate (Forest Tree Breeding, Genetics and Physiology Unit UAGPF, INRA Orléans) for providing poplar samples and B.Chabbert (Joint Research Unit for Fractionation of Agricultural Resources and Packaging FARE, INRA Reims), who carried out bio-chemical analysis.

# Figures

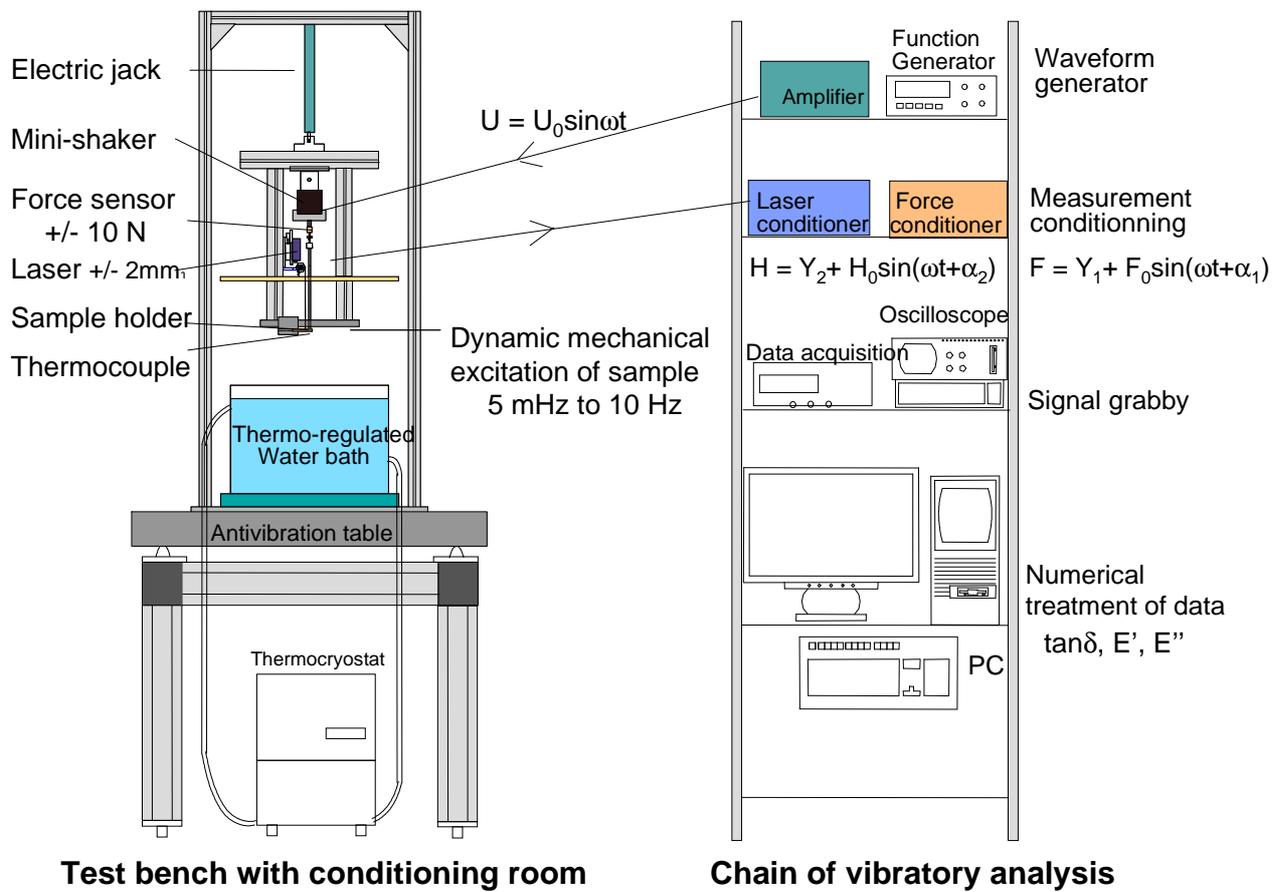

**Figure 1: Schematic representation of the experimental device used: the WAVE$^T$**

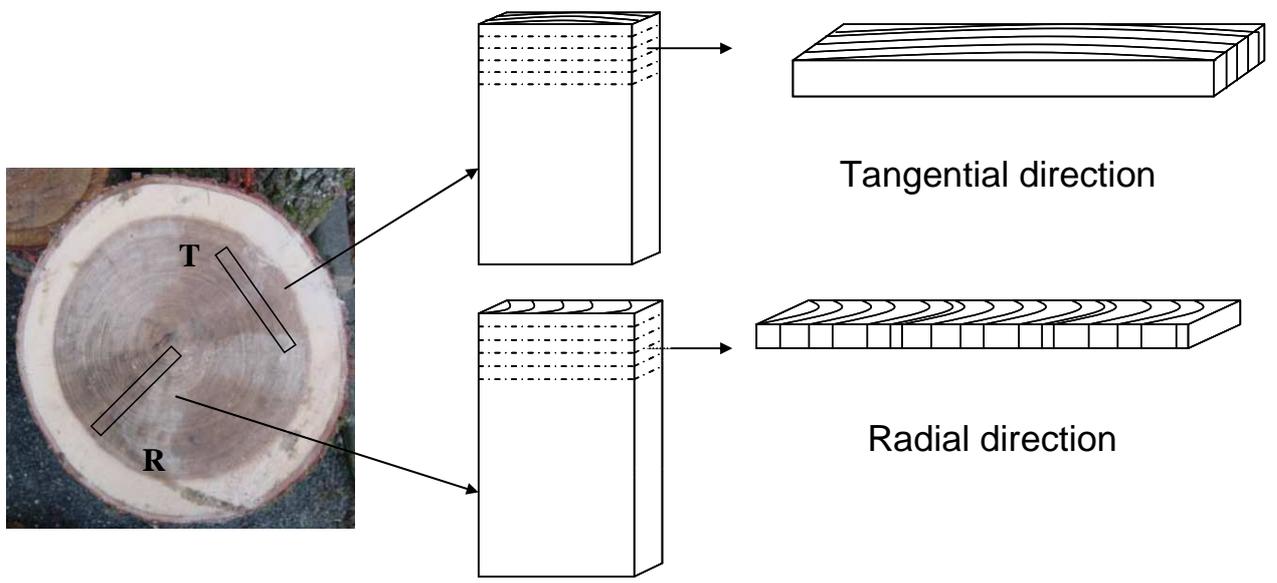

**Figure 2: Wood sampling**

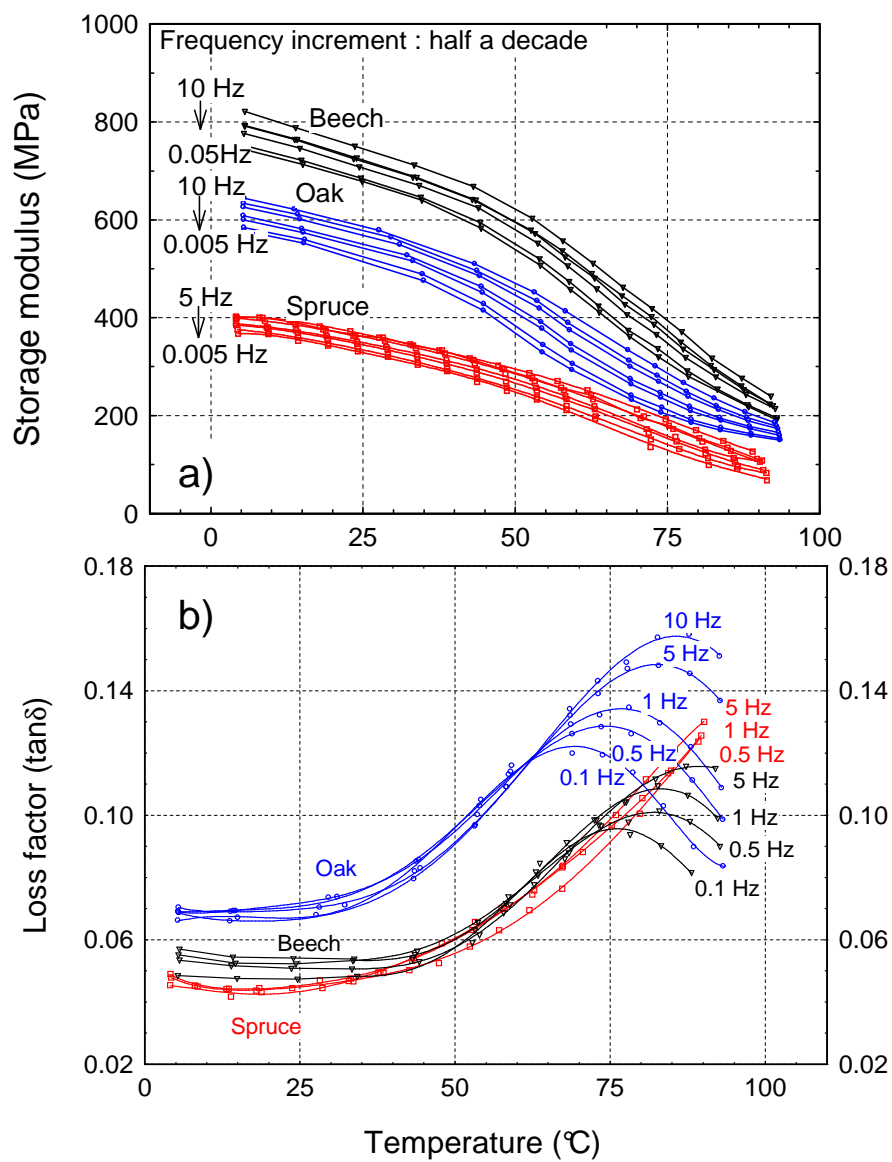

**Figure 3: Evolution of the viscoelastic properties versus temperature of Oak, Beech and Spruce samples. Normal wood, radial direction. a – Storage Modulus. b – Loss factor.**

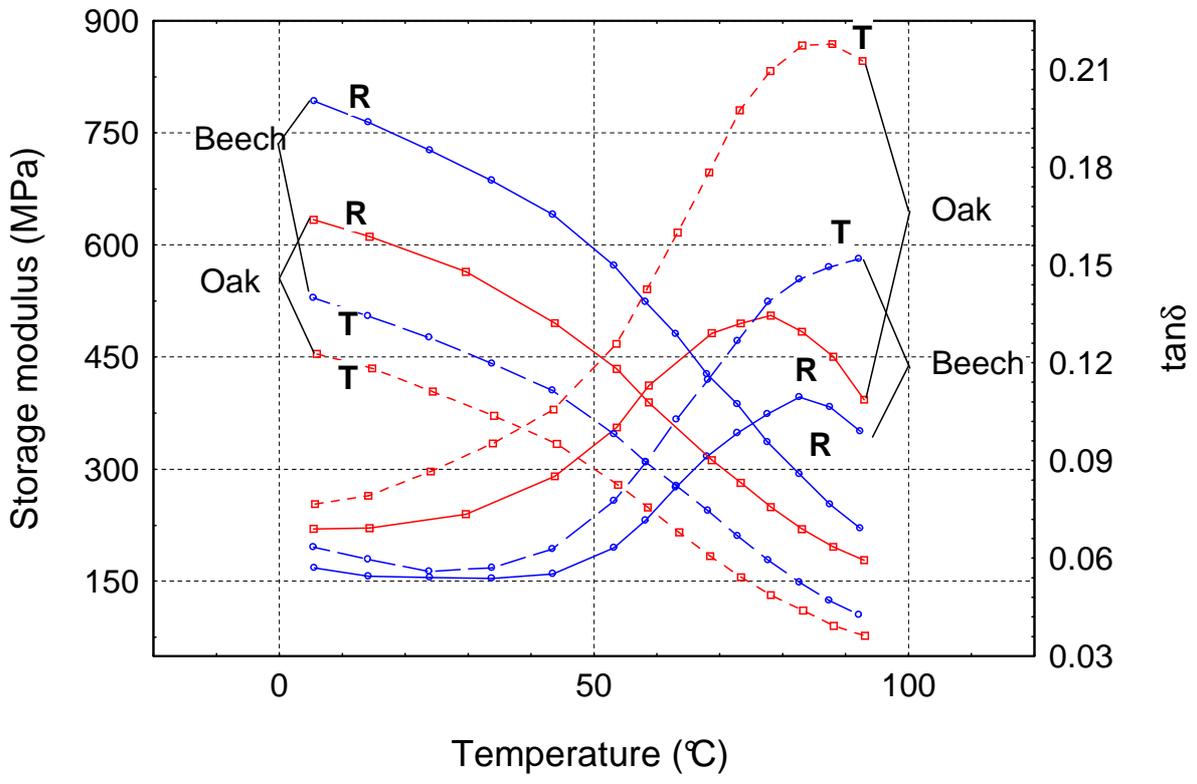

**Figure 4: Evolution of the storage modulus and the loss factor versus temperature of oak and beech samples in radial and tangential direction (frequency: 1 Hz)**

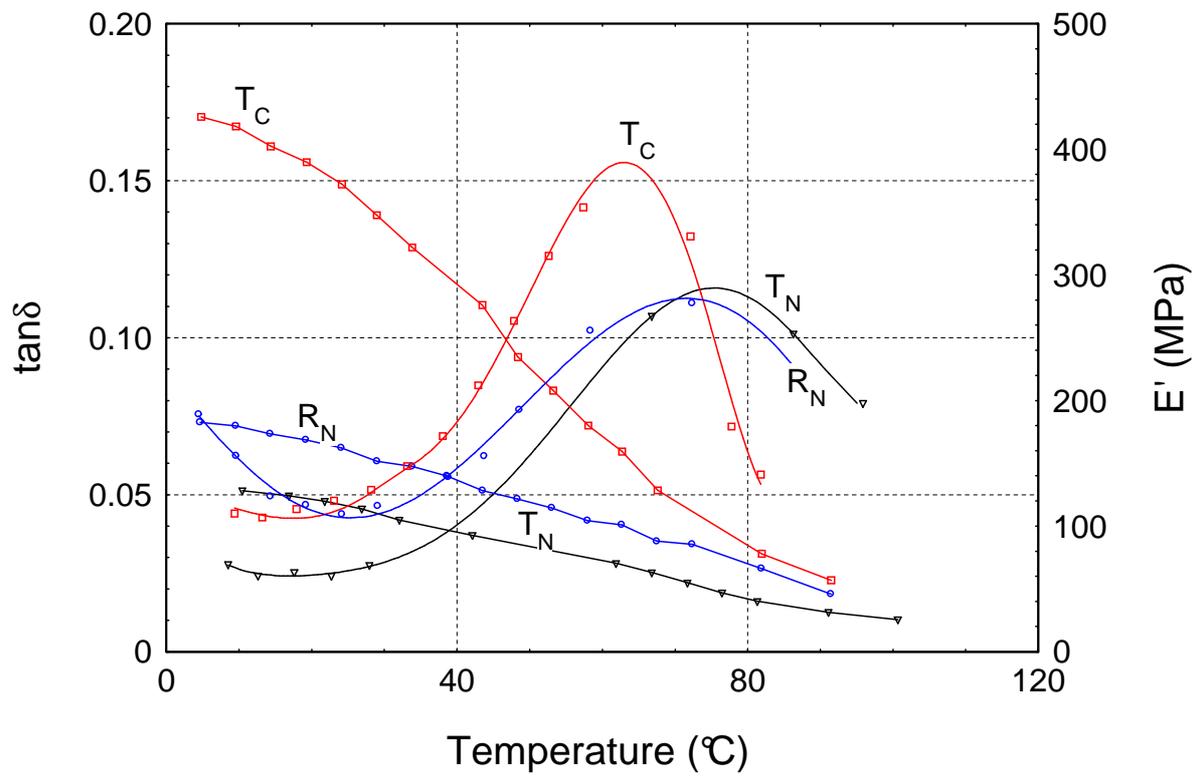

**Figure 5: Evolution of the storage modulus and the loss factor versus temperature of normal wood and compression wood of Spruce (frequency: 5 mHz)**

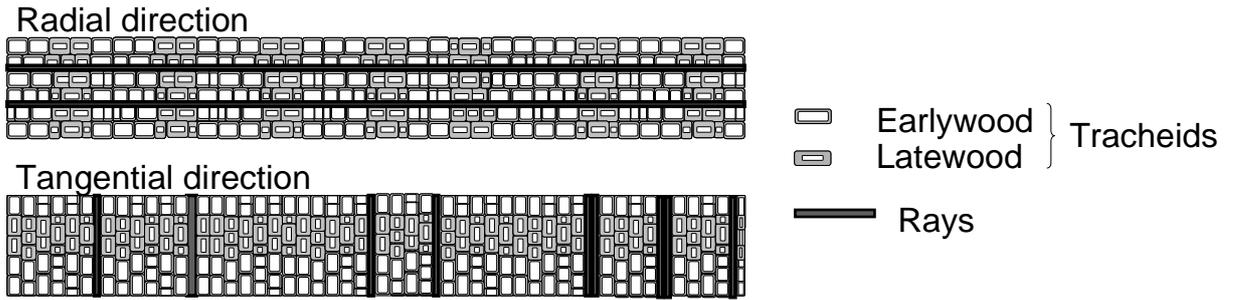

*Wood behaviour in flexion tests according to the material direction*

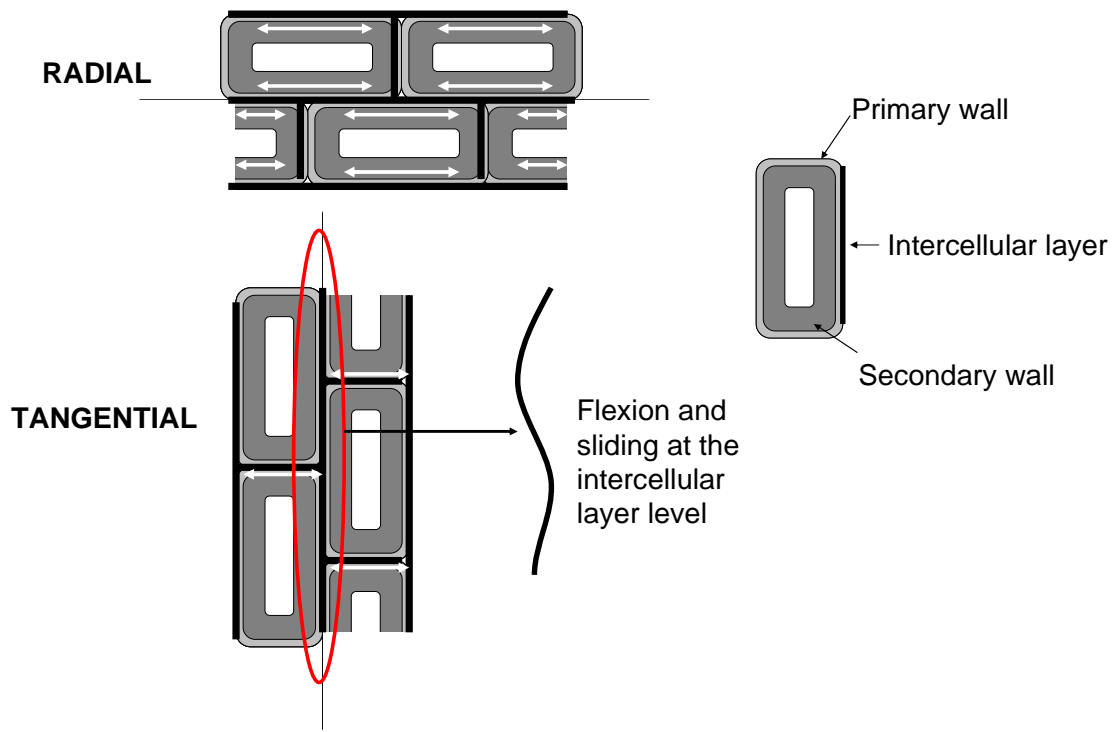

**Figure 6: Schematic representation of softwood anatomical organisation in radial and tangential direction.**

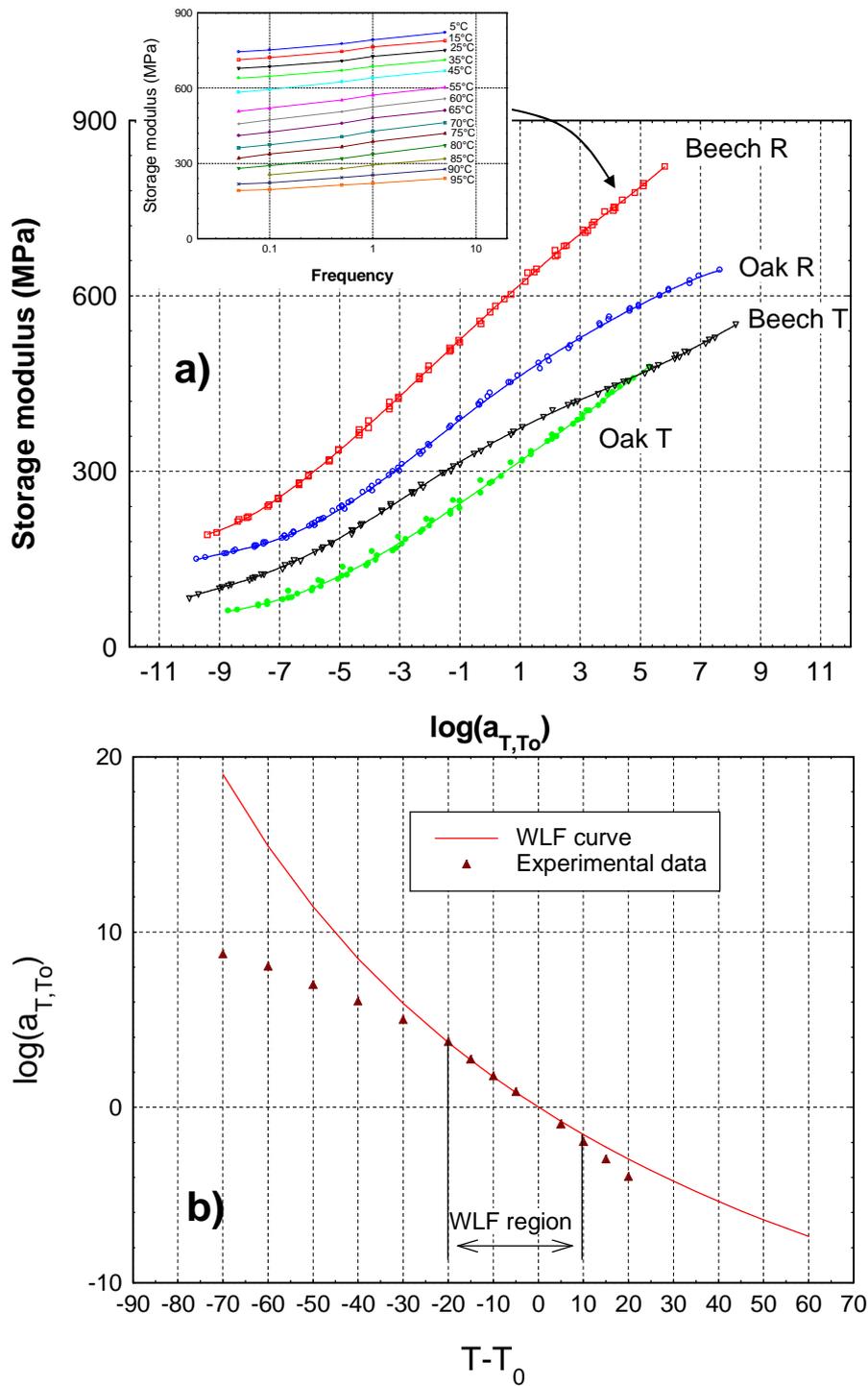

**Figure 7 -a: Master curves for beech and oak wood in radial and tangential directions.  b – The shift factor versus the temperature T-T$_0$ for the master curve of beech in radial direction. (Temperature of reference: 75°C).**

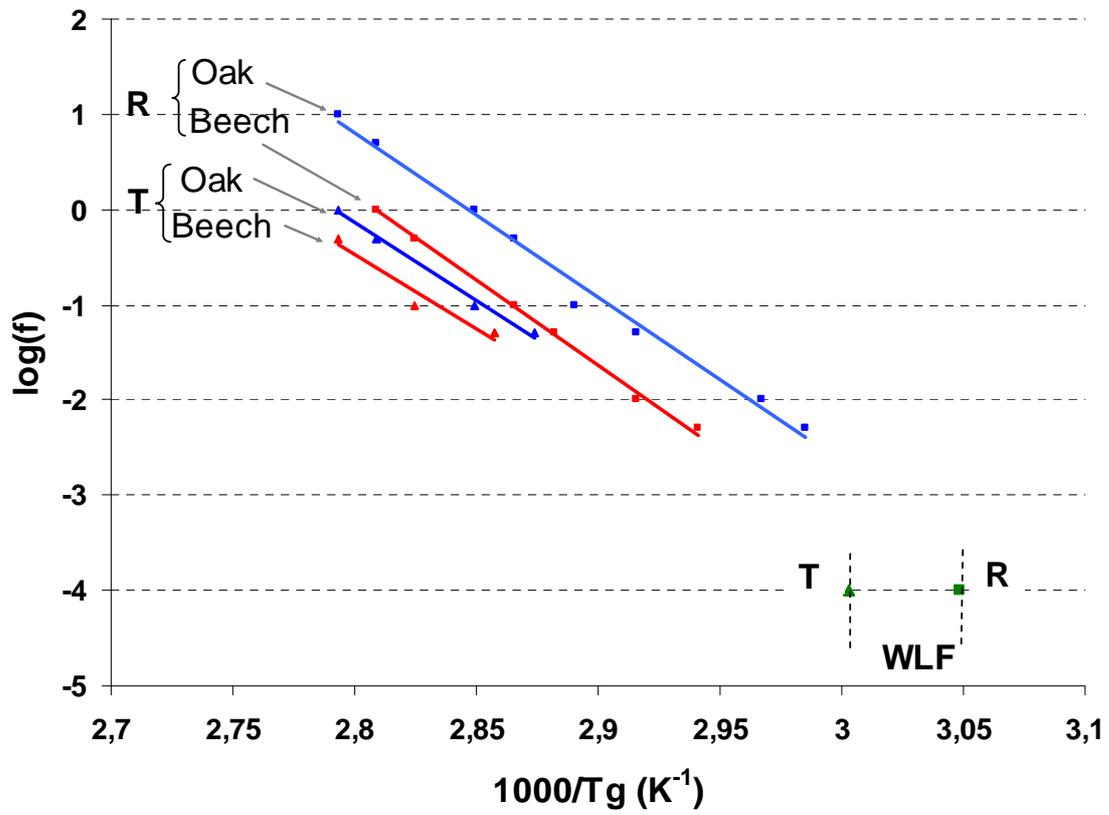

**Figure 8: Arrhenius plot for beech and oak wood in radial and tangential directions**

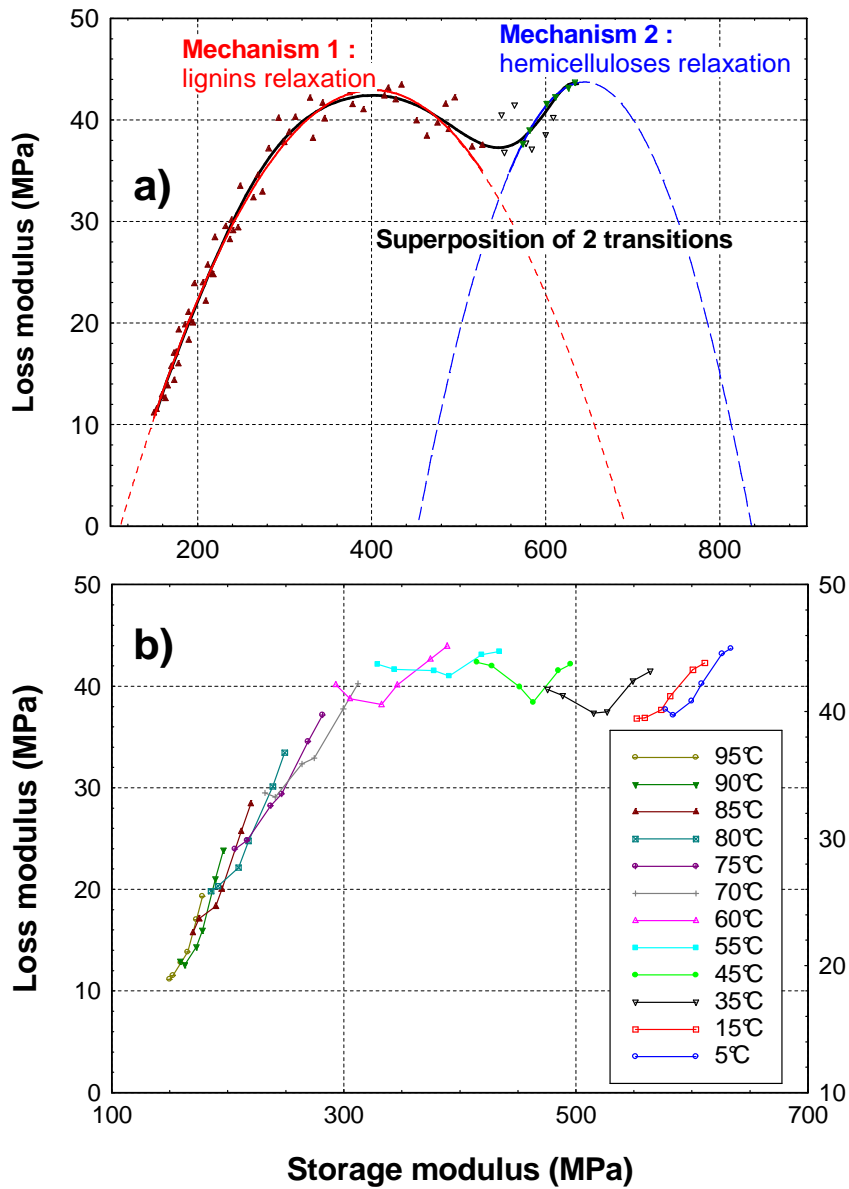

**Figure 9- a : Cole-Cole plot for oak in radial direction. Measurements were carried out in the frequency range of $5.10^{-3}$ Hz to 1 Hz and for temperatures between 5 and 95°C.   b – Same experimental data than in graph a, plotted according to the temperature.**

# Abbreviations:

$R_{NW}$: Radial direction, Normal Wood

$T_{NW}$: Tangential direction, Normal Wood

$T_{CW}$: Tangential direction, Compression Wood

$R_{OW}$: Radial direction, Opposite Wood

$T_{OW}$: Tangential direction, Opposite Wood

$R_{TW}$: Radial direction, Tension Wood

$T_{TW}$: Tangential direction, Tension Wood

# Tables

**Table 1: Summary of the viscoelastic properties of tested wood**

| Specie | Oak | | Beech | | Spruce | | | Poplar | | | |
|---|---|---|---|---|---|---|---|---|---|---|---|
| Material direction | R | T | R | T | $R_{NW}$ | $T_{CW}$ | $T_{NW}$ | $R_{OW}$ | $R_{TW}$ | $T_{OW}$ | $T_{TW}$ |
| Number of tested samples | 3 | 3 | 5 | 4 | 5 | 10 | 10 | 1 | 1 | 3 | 3 |
| Infradensity (kg.m$^{-3}$) | | | 585 (575…595) | 592 (591…596) | 290 (287…309) | 520 (502…549) | 290 (282…297) | | | | |
| Standard deviation | | | 7 | 3 | 11 | 16 | 9 | | | | |
| E' (20°C, 1Hz) (MPa) | 645 (600…728) | 457 (420…528) | 674 (583…779) | 491 (472…504) | 213 (185…245) | 447 (418…485) | 84 (81…87) | 680 | 495 | 242 (210…275) | 146 (139…153) |
| Standard deviation | 72 | 62 | 84 | 14 | 39 | 34 | 3 | | | 46 | 10 |
| $E'_R/E'_T$ | | 1,41 | | 1,37 | | | 2,54 | | | 2,81 | 3,39 |
| Tg (°C) | | | | | | | | | | | |
| 10 Hz | 85 | | | | | | | | | | |
| 5 Hz | 83 | | | | | | | | 85 | | |
| 1 Hz | 78 | 85 | 83 | | | | | | 81 | | |
| 0.5 Hz | 76 | 83 | 81 | 85 | | | | | 79 | | |
| 0.1 Hz | 73 | 78 | 76 | 81 | | | | 78 | 74 | 85 | 76 |
| 0.05 Hz | 70 | 75 | 74 | 77 | | | | | | | |
| 0.01 Hz | 64 | | | | | | | | | | |
| 0.005 Hz | 62 | | | | 70 | 61 | 75 | | | | |

**Table 2: Values of the storage modulus across the grain at 5 and 95°C for different wood species**

| | | E' (1 Hz, 5°C) (MPa) | E' (1 Hz, 90°C) (MPa) | ratio | Decreasing (%) |
|---|---|---|---|---|---|
| Oak | R | 630 | 185 | 3,4 | -71% |
| | T | 455 | 85 | 5,4 | -81% |
| Beech | R | 790 | 235 | 3,4 | -70% |
| | T | 530 | 115 | 4,6 | -78% |
| Poplar | $R_{TW}$ | 530 | 145 | 3,7 | -73% |
| | $R_{OW}$ | 705 | 240 | 2,9 | -66% |
| | $T_{TW}$ | 170 | 40 | 4,3 | -76% |
| | $T_{OW}$ | 235 | 55 | 4,3 | -77% |
| Spruce | $R_{NW}$ | 200 | 55 | 3,6 | -73% |
| | $T_{CW}$ | 485 | 85 | 5,7 | -82% |
| | $T_{NW}$ | 95 | 20 | 4,8 | -79% |

**Table 3: Characteristics of the WLF and Arrhenius laws for oak and beech samples**

(For the WLF law, the reference temperature chosen corresponds to the softening temperature of wood at very low frequency, identify using the graph of $(T-T_0)/\log a_T$ versus $(T-T_0)$).

|  | Specie | Oak | | Beech | |
| --- | --- | --- | --- | --- | --- |
|  | Material direction | R | T | R | T |
|  | Arrhénius $\Delta Ha$ (kJ.mol$^{-1}$) | 332 | 314 | 341 | 299 |
| WLF | T$_{réf}$ = T$_g$ | 55°C | 60°C | 55°C | 60°C |
| WLF | C$_1$ | 24,5 | 27,4 | 23,3 | 24,9 |
| WLF | C$_2$ | 111,9 | 132,20 | 109 | 97,70 |
| WLF | $\Delta Ha$ (kJ.mol$^{-1}$) | 426 | 440 | 453 | 490 |